\documentstyle[12pt]{article}

\def\hybrid{\topmargin -20pt  \oddsidemargin 0pt
      \headheight 0pt   \headsep 0pt
      \textwidth 6.25in % A4 paper
      \textheight 9.5in % A4 paper
      \marginparwidth .875in
      \parskip 5pt plus 1pt   \jot = 1.5ex}

\hybrid

\begin{document}
%\titlepage
\def\x{\times}
\def\beq{\begin{equation}}
\def\eeq{\end{equation}}
\def\beqa{\begin{eqnarray}}
\def\eeqa{\end{eqnarray}}
\def\cD{ {\cal D}}
\def\L{ {\cal L}}
\def\C{ {\cal C}}
\def\N{ {\cal N}}
\def\calE{{\cal E}}
\def\lin{{\rm lin}}
\def\Tr{{\rm Tr}}
\def\mxth{\mathsurround=0pt }
\def\xversim#1#2{\lower2.pt\vbox{\baselineskip0pt \lineskip-.5pt
x  \ialign{$\mxth#1\hfil##\hfil$\crcr#2\crcr\sim\crcr}}}
\def\simgr{\mathrel{\mathpalette\xversim >}}
\def\simle{\mathrel{\mathpalette\xversim <}}

\renewcommand{\a}{\alpha}
\renewcommand{\b}{\beta}
\renewcommand{\c}{\gamma}
\renewcommand{\d}{\delta}
\newcommand{\th}{\theta}
\newcommand{\TH}{\Theta}
\newcommand{\pa}{\partial}
\newcommand{\g}{\gamma}
\newcommand{\G}{\Gamma}
\newcommand{\A}{\Alpha}
\newcommand{\B}{\Beta}
\newcommand{\D}{\Delta}
\newcommand{\e}{\epsilon}
\newcommand{\E}{\Epsilon}
\newcommand{\z}{\zeta}
\newcommand{\Z}{\Zeta}
\newcommand{\k}{\kappa}
\newcommand{\K}{\Kappa}
\renewcommand{\l}{\lambda}
\renewcommand{\L}{\Lambda}
\newcommand{\m}{\mu}
\newcommand{\M}{\Mu}
\newcommand{\n}{\nu}
\newcommand{\X}{\Chi}
\newcommand{\R}{\Rho}
\newcommand{\s}{\sigma}
\renewcommand{\S}{\Sigma}
\renewcommand{\t}{\tau}
\newcommand{\T}{\Tau}
\newcommand{\y}{\upsilon}
\newcommand{\Y}{\upsilon}
\renewcommand{\o}{\omega}
\newcommand{\q}{\theta}
\newcommand{\h}{\eta}

\def\dota{ {\dot{\alpha}} }
\def\lag{Lagrangian}
\def\Kahler{K\"{a}hler}
\def\kahler{K\"{a}hler}
\def\A{ {\cal A}}
\def\C{ {\cal C}}
\def\D{ {\cal D}}
\def\F{{\cal F}}
\def\L{ {\cal L}}

\def\R{ {\cal R}}
\def\x{ \times }
\def\beq{\begin{equation}}
\def\eeq{\end{equation}}
\def\beqa{\begin{eqnarray}}
\def\eeqa{\end{eqnarray}}

\sloppy
\newcommand{\be}{\begin{equation}}
\newcommand{\eq}{\end{equation}}
\newcommand{\ov}{\overline}
\newcommand{\un}{\underline}
\newcommand{\p}{\partial}
\newcommand{\la}{\langle}
\newcommand{\ra}{\rangle}
\newcommand{\bl}{\boldmath}
\newcommand{\ds}{\displaystyle}
\newcommand{\nl}{\newline}
\newcommand{\Nzahl}{{\bf N}  }
\newcommand{\zzahl}{ {\bf Z} }
\newcommand{\Zzahl}{ {\bf Z} }
\newcommand{\Qzahl}{ {\bf Q}  }
\newcommand{\Rzahl}{ {\bf R} }
\newcommand{\Czahl}{ {\bf C} }
\newcommand{\wt}{\widetilde}
\newcommand{\wh}{\widehat}
\newcommand{\fs}[1]{\mbox{\scriptsize \bf #1}}
\newcommand{\ft}[1]{\mbox{\tiny \bf #1}}
\newtheorem{satz}{Satz}[section]
\newenvironment{Satz}{\begin{satz} \sf}{\end{satz}}
\newtheorem{definition}{Definition}[section]
\newenvironment{Definition}{\begin{definition} \rm}{\end{definition}}
\newtheorem{bem}{Bemerkung}
\newenvironment{Bem}{\begin{bem} \rm}{\end{bem}}
\newtheorem{bsp}{Beispiel}
\newenvironment{Bsp}{\begin{bsp} \rm}{\end{bsp}}
\renewcommand{\arraystretch}{1.5}

%\textwidth14.5cm
%\textheight23.0cm
%\oddsidemargin0.5cm
%\topmargin-1.4cm

%\addtocounter{section}{1}

\renewcommand{\thesection}{\arabic{section}}
\renewcommand{\theequation}{\thesection.\arabic{equation}}

%\setcounter{section}{1}
%\addtocounter{section}{1}
\parindent0em

\def\S4{\frac{SO(4,2)}{SO(4) \otimes SO(2)}}
\def\P3{\frac{SO(3,2)}{SO(3) \otimes SO(2)}}
\def\MGd{\frac{SO(r,p)}{SO(r) \otimes SO(p)}}
\def\SOd{\frac{SO(r,2)}{SO(r) \otimes SO(2)}}
\def\SO2{\frac{SO(2,2)}{SO(2) \otimes SO(2)}}
\def\SUm{\frac{SU(n,m)}{SU(n) \otimes SU(m) \otimes U(1)}}
\def\SUS{\frac{SU(n,1)}{SU(n) \otimes U(1)}}
\def\SK{\frac{SU(2,1)}{SU(2) \otimes U(1)}}
\def\SU{\frac{ SU(1,1)}{U(1)}}

%%%%%%%%%%%%%%%%%%%%%%%%%%%%%%%%%%%%%%%%%%%%%%%%%%%%%

\begin{titlepage}
\begin{center}
\hfill CERN-TH/96-223\\
\hfill HUB-EP-96/42\\
\hfill {\tt hep-th/9608099}\\

\vskip .1in

{\bf MODULAR SYMMETRIES OF\\ $N=2$ BLACK HOLES}

\vskip .2in

{\bf Gabriel Lopes Cardoso$^a$, 
Dieter L\"ust$^b$ and Thomas Mohaupt$^b$}\footnote{ 
Email: \tt cardoso@surya11.cern.ch, luest@qft1.physik.hu-berlin.de, \hfill
\\
mohaupt@qft2.physik.hu-berlin.de}
\\
\vskip 1.2cm

$^a${\em Theory Division, CERN, CH-1211 Geneva 23, Switzerland}\\
$^b${\em Humboldt-Universit\"at zu Berlin,
Institut f\"ur Physik, 
D-10115 Berlin, Germany}\\

\vskip .1in

\end{center}

\vskip .2in

\begin{center} {\bf ABSTRACT } \end{center}
\begin{quotation}\noindent
We discuss the transformation properties of classical
extremal $N=2$ black hole
solutions in $S$-$T$-$U$ like models under $S$ and $T$ duality.
Using invariants of (subgroups of) the triality group, which
is the symmetry group
of the classical BPS mass formula, the transformation properties
of the moduli on the event horizon and of the entropy under 
these transformations become manifest. We also comment on quantum corrections
and we make a conjecture
for the one-loop corrected entropy.

\end{quotation}
\vskip 5cm
August 1996\\
\hfill CERN-TH/96-223 \\
\end{titlepage}
\vfill
\eject

%%%%%%%%%%%%%%%%%%%%%%%%%%%%%%%%%%%%%%%%%%%%%%%%%%%%%%%%%%%%%
\newpage

\section{Introduction}

\setcounter{footnote}{0}

During the last years it has become obvious that the structure
behind string theories is organized by discrete perturbative 
and non--perturbative transformations, which are either symmetries
of a given string theory or map two different theories into one another.
(See \cite{DualRev} for recent reviews and references.)

The pattern of relations that arises this way becomes more and more
complicated when the number $D$ of space--time dimensions and
the number $N$ of supersymmetries is decreased.

The case of $D=4, N=2$ models has proved to be especially interesting
because it is rich in structure, but still exactly tractable.
The perturbative aspects and the role of symplectic transformations
were worked out in \cite{CDFP}, \cite{DKLL}, \cite{AFGNT}, \cite{HarMor}.
One of the most prominent examples in this class is the so--called
$S$-$T$-$U$ model, which has just the minimal number 3 of vector
multiplets for a theory coming from a $D=6, N=1$ model by
toroidal compactification.\footnote{The model has in addition a large number
of hypermultiplets which are not relevant for our purpose.}

These vector multiplets contain the dilaton/axion $S$ and the two
moduli of the torus, $T$ and $U$ as their scalar components.

The $S$-$T$-$U$ model can be obtained from the ten--dimensional
heterotic $E_8 \times E_8$ string
by compactification on $K3 \times T_2$ with instanton
numbers (14,10). It is related by well
established non--perturbative dualities to the $IIA$ string
compactified on the Calabi--Yau--threefold ${\bf P}_{1,1,2,8,12}(24)$
\cite{KacVaf}
and to the $IIB$ string on the mirror, as well as to the type $I$ superstring 
by a more recently 
proposed duality \cite{ABFPT}. Moreover, 
there are self--dualities under perturbative and non--perturbative 
transformations which can be derived either from the triality
of $D=4,N=4$ heterotic, $IIA$ and $IIB$ models \cite{DufLiuRah}
or from the
self--duality of the corresponding heterotic $D=6,N=1$ model \cite{DufMinWit}.

The self--dualities manifest themselves very clearly in the spectrum
of BPS states, which is organized by the symplectic structure
of local $N=2$ supersymmetry. The BPS spectrum consists
of both elementary and solitonic states, where the latter ones can
be explicitly constructed as extremal black hole solutions of the low
energy effective action. (See \cite{Tse} for a review on stringy
extremal black holes.)
Extremal $D=4, N=2$ black holes with a 
non--vanishing
event horizon have 
the property 
that the moduli take on the horizon certain fixed point
values, which solve an extremization problem for the central
charge of the $N=2$ algebra
\cite{ExtrN2BHs}, \cite{FerKal1}.
(This seems to have generalizations for theories
with higher $D$ and $N$ \cite{AndDAuFer}.)
Moreover the absolute
value squared of the central charge coincides with the 
extremized ADM mass squared of the extremal black hole and - 
up to a constant - with its entropy.
All these quantities can be expressed in terms of the symplectic
quantum numbers of the solution.

The BPS spectrum of the theory receives both perturbative (one-loop)
and non--perturbative corrections whose form is restricted
by the symplectic structure of local $N=2$ supersymmetry. In this note
we will focus mostly on duality properties of the classical BPS 
spectrum, i.e. we consider both elementary and solitonic states
but ignore quantum corrections.  

As discussed in \cite{DufLiuRah} and in \cite{CCLMR},
the classical BPS mass formula\footnote{Note that this does not
imply the existence of all the corresponding states. In fact
the full quantum
spectrum is not expected to have triality symmetry. See section
\ref{quantum} for a discussion of quantum properties.}
of the $S$-$T$-$U$ model is invariant under 
the triality group
\be
(SL(2, {\bf Z})_S \otimes SL(2, {\bf Z})_T \otimes
SL(2, {\bf Z})_U ) \times {\bf Z}_2^{T-U}\times {\bf Z}_2^{S-T}
\times {\bf Z}_2^{S-U} \;\;,
\eq
where the $SL(2,{\bf Z})$ factors act as fractional linear transformations
and the ${\bf Z}_2$ factors act as permutations on the moduli $S,T,U$.
The spectrum decomposes into certain subsets called orbits, which
can be characterized by invariants of the triality group and of
certain subgroups as discussed in \cite{CCLMR}. The purpose of this
note is to use the formalism developed in \cite{CCLMR} to make
explicit the symmetry and transformation properties of extremal
black hole solutions of the $S$-$T$-$U$ model. Note that
every $D=4,N=2$ model coming from $D=6$ by toroidal compactification
will contain the solitons discussed here as a subset. Thus we are
discussing the universal sector of all these $S$-$T$-$U$ like models.

The paper is organized as follows: In section \ref{solution} the conditions
for extremality of the central charge are solved using invariants
of subgroups of the triality group. In section \ref{classical} the entropy is 
computed and found to be completely triality invariant. 
We also comment on the relation to $D=4, N=4$ models.
The final sections \ref{quantum} contains some remarks 
on quantum corrections. 
By considering the transformation properties of the classical
entropy under one--loop $T$ duality we arrive at a conjecture for 
the one--loop corrected entropy. We also speculate on what happens at the 
non--perturbative level.

\section{The solution on the horizon \label{solution}}
\setcounter{equation}{0}

%%%%%%%%%%%%%%%%%%%%%%%%%%%%%%%%%%%%%%%%%%%%%%%%%%%%%%%%%%%%%%%

Let us first recall some relevant elements of the symplectic formalism
of $N=2$ supergravity in the concrete case of supergravity coupled
to $n_V$ vectormultiplets. We follow
references \cite{CDFP}, \cite{DKLL}, \cite{AFGNT}, \cite{CaLuMo},
which can be consulted for more complete information.
The low energy effective action of $N=2$ supergravity coupled to
$n_V$ vectormultiplets can be written in terms of a so--called
symplectic section $\Omega^T = (P^I, iQ_I)$, $I=0,\ldots,n_V$.
The combined set of field equations and Bianchi identities is
invariant under symplectic transformations $\Gamma \in Sp(2(n_v+1))$,
which act on the section $\Omega$ as
\be
\left( \begin{array}{c}
P^I \\ iQ_I \\
\end{array} \right) 
\rightarrow
\Gamma 
\left( \begin{array}{c}
P^I \\ iQ_I \\
\end{array} \right) = 
\left( \begin{array}{cc}
U & Z \\ W & V \\
\end{array} \right)
\left( \begin{array}{c}
P^I \\ iQ_I \\
\end{array} \right) \;\;.
\label{SympTrans}
\eq
Whereas at the classical level the symplectic transformations can
be continuous, $\Gamma \in Sp(2(n_V+1),{\bf R})$, it is expected that
this is broken to a discrete subgroup by instanton effects at the 
quantum level, $\Gamma \in Sp(2(n_V+1),{\bf Z})$.

The mass formula for BPS saturated states is given by
\be 
M_{BPS}^2 =|z|^2 = |M_I P^I + i N^I Q_I|^2 \;\;,
\eq
where $z$
is the central charge of the $N=2$ 
supersymmetry algebra and  
$M_I$ and $N^I$ are the symplectic
quantum numbers. Note that the $M_I$ are related to electric and the 
$N^I$ to magnetic charges under the $U(1)^{n_V+1}$ gauge 
group.\footnote{The extra $U(1)$ corresponds to the graviphoton.}
The BPS mass is invariant under symplectic transformations
(\ref{SympTrans}) provided the quantum numbers are redefined
by $(N^I , -M_I) \rightarrow (N^I,-M_I) \Gamma^T$.

The $2(n_V+1)$ components of $\Omega$ 
can be expressed in terms of the $n_V$ physical scalar fields,
which provide so--called special coordinates on the moduli space.
In the case of the $S$-$T$-$U$ model there are three such scalars,
namely the dilaton $S$ and the moduli $T$ and $U$. One choice for
$\Omega$ at the classical level is given by
\be
\Omega^T = (P^I, iQ_I) = e^{K/2}(1, TU, iT, iU, iSTU, iS, -SU, -ST) \;\;,
\label{PQsection}
\eq
where 
\be
K = - \log(S+\ov{S})(T+\ov{T})(U+\ov{U}) 
\eq
is the K\"ahler potential. 
The BPS mass formula then takes the form
\be
M^2_{BPS} = |z|^2 = e^K | {\cal M} |^2  \;\;,
\eq
where
\be
{\cal M} = M_0 + M_1 TU + i M_2 T + i M_3 U + i N^0 STU + i N^1 S
- N^2 SU - N^3 ST 
\eq
is the so--called holomorphic mass.

Symmetry transformations in $N=2$ supergravity coupled to
vector multiplets must act as $Sp(2(n_V+1), {\bf Z})$
transformations. When acting on the section
(\ref{PQsection}), symplectic transformations with
$W=Z=0$ (implying $V = U^{T,-1}$) leave the action invariant,
whereas transformations with $Z=0$ leave it invariant up to
total derivatives. This is the form classical and perturbative
symmetries must take. On the other hand transformations with
$Z\not=0$ are not symmetries of the action and exchange
electric and magnetic degrees of freedom, which is the
suitable form for non--perturbative symmetries or dualities.

The maximal known symmetry group of
the $S$-$T$-$U$ model at the classical level
is the triality group \cite{DufLiuRah}, \cite{CCLMR}
\be
(SL(2, {\bf Z})_S \otimes SL(2, {\bf Z})_T \otimes
SL(2, {\bf Z})_U ) \times {\bf Z}_2^{T-U}\times {\bf Z}_2^{S-T}
\times {\bf Z}_2^{S-U} \;\;.
\label{Triality}
\eq
It contains the tree level $T$ duality group
$O(2,2,{\bf Z})_{T,U} \sim (SL(2,{\bf Z})_T \otimes
SL(2,{\bf Z})_U) \times {\bf Z}_2^{T-U}$, 
which is a classical symmetry,
together with
the $S$ duality group $SL(2,{\bf Z})_S$ \cite{FILQ}
and the
exchange transformations $S \leftrightarrow T$ and 
$S \leftrightarrow U$, which are
non-perturbative transformations.\footnote{Since 
the terminology might be confusing, let us recall
that in this context discussing the theory at 'classical' or 
'semi--classical' level means that one includes both elementary and
solitonic states, but that one ignores all quantum corrections.
'Classical' transformations then leave the action invariant,
whereas 'non--perturbative' transformations map it to a dual
action which (in the case under consideration ) has the same form, but
contains a different set of degrees of freedom as the elementary ones.}

As a concrete example let us specify the symplectic matrices
realizing the $S$ duality transformations
$S \rightarrow \frac{aS -ib}{icS + d}$, $T \rightarrow T$,
$U \rightarrow U$:
\be
U = d {\bf 1},\;\;\;V = a {\bf 1},\;\;\;W = b {\bf H},\;\;\;
Z = c {\bf H} \;\;,
\eq
where ${\bf H} = \eta \oplus \eta$ and $\eta = \left(
\begin{array}{cc} 0&1\\1&0\\ \end{array} \right)$.
Also note that the action of the exchange transformations
will be given below in (\ref{Exchange}), whereas the
matrices of $O(2,2,{\bf Z})_{T,U}$ transformations can be found
in \cite{CaLuMo}.

Let us then consider classical BPS black hole solutions of $S$-$T$-$U$ like
$N=2$ supergravity coupled to (at least) 3 vector multiplets with
four electric and four magnetic charges $M_0, \ldots, N^3$. 
In \cite{FerKal1} it was argued that the central charge becomes
extremal on the horizon. This can be used to express the moduli
$S,T,U$ on the horizon in terms of the quantum number 
$M_I, N^I$. 
Moreover, the entropy ${\cal S}$ of the black hole is proportional
to the absolute value squared of the
extremized central charge $z_{hor}$ and to the extremized ADM mass
$M_{ADM}$ of the black hole:
\be
{\cal S} = \pi M_{ADM}^2,\;\;\;
M_{ADM}^2 = |z_{hor}|^2 = e^K | {\cal M}_{hor} |^2  \;\;.
\eq
To be precise, the central charge $z$ has to be extremized with 
respect to the K\"ahler covariant derivative \cite{FerKal1}, that
is by solving\footnote{
The lower indices indicate ordinary partial derivatives with
respect to the special coordinates $S,T,U$.}
\beqa
D_i z = 0 \;\;  \rightarrow \;\; 
e^{K/2} \left(K_i {\cal M} + {\cal M}_i \right) =0 \;\;
, \;\; i = S, T, U  \;\;.
\eeqa
This yields three quadratic equation for the three moduli $S,T,U$:
\begin{eqnarray}
\ov{S} &=& \frac{
M_0 + M_1 TU + i M_2 T + i M_3 U}{
i N^0 TU + i N^1 - N^2 U - N^3 T} \;\;, \nonumber \\
\ov{T} &=& \frac{
M_0 + i M_3 U + i N^1 S - N^2 SU}{
M_1 U + i M_2 + i N^0 SU - N^3 S} \;\;, \nonumber \\
\ov{U} &=& \frac{
M_0 + i M_2 T + i N^1 S - N^3 ST}{
M_1 T + i M_3 + i N^0 ST - N^2 S} \;\;.
\label{QuadEq}
\end{eqnarray}
The equations are related by the exchange transformations
\be
\begin{array}{llll}
{\bf Z}_2^{S-T}: &
S \leftrightarrow T, &
M_2 \leftrightarrow N^1, &
M_1 \leftrightarrow -N^2 \;\;, \\
{\bf Z}_2^{S-U}: &
S \leftrightarrow U, &
M_3 \leftrightarrow N^1, &
M_1 \leftrightarrow -N^3 \;\;, \\
{\bf Z}_2^{T-U}: &
T \leftrightarrow U,&
M_2 \leftrightarrow M_3,&
N^2 \leftrightarrow N^3  \;\;.\\
\end{array}
\label{Exchange}
\eq
Thus, it is natural to expect that the extremized moduli as well as
the entropy have simple and natural transformation (or invariance)
properties under $S$ and $T$ duality transformations. As discussed above 
(the relevant  subsector of) 
the tree level BPS mass formula of $S$-$T$-$U$ like
$N=2$ models is invariant under the triality group
(\ref{Triality}).
In order
to make transformation properties manifest, one can use the invariants
of certain subgroups of the triality groups, as discussed 
in appendix A of \cite{CCLMR}. 
For example, taking mutual $O(2,2)$ scalar products
\be
\la v, w \ra = v_0 w_1 + v_1 w_0 + v_2 w_3 + v_3 w_2
\eq 
of the vectors
$M = (M_0, \ldots, M_3)$ and $N = (N^0,\ldots, N^3)$ gives rise
to the invariants
\begin{eqnarray} 
\la M, M \ra &=& 2 M_0 M_1 + 2 M_2 M_3 \nonumber\\
\la N, N \ra &=& 2 N^0 N^1 + 2 N^2 N^3  \nonumber\\
\la M, N \ra &=&  M_0 N^1 + M_1 N^0 +  M_2 N^3 + M_3 N^2 
\end{eqnarray}
of the $T$ duality group
\be
O(2,2,{\bf Z})_{(T,U)} = ( SL(2, {\bf Z})_T \otimes SL(2, {\bf Z})_U )
\times {\bf Z}_2^{T-U} \;\;.
\eq
The exchange symmetries $S\leftrightarrow T$ and 
$S \leftrightarrow U$ map the vectors $M,N$ to
vectors $M',N'$ and $M'', N''$ whose components can be read off from
(\ref{Exchange}). Using these vectors one obtains
the invariants $\la M',M'\ra$, etc. 
and $\la M'',M'' \ra$, etc. of the groups
$O(2,2,{\bf Z})_{S,U} \sim (SL(2,{\bf Z})_S \otimes SL(2,{\bf Z})_U)
\times {\bf Z}_2^{S-U}$ 
and $O(2,2,{\bf Z})_{S,T} \sim (SL(2,{\bf Z})_S \otimes SL(2,{\bf Z})_T)
\times {\bf Z}_2^{S-U}$. These groups are the classical
symmetries of the triality rotated actions, where $T$ or $U$
have taken over the role of the dilaton.

Moreover, there is a $GL(4)$ subgroup of $Sp(8)$ defined
by $W=Z=0$ in (\ref{SympTrans}). In this subgroup $V = U^{T,-1}$
and therefore 
\be
M \cdot N = M_0 N^0 + \cdots + M_3 N^3
\eq
is invariant.  Since the tree level $T$ duality group 
$O(2,2,{\bf Z})_{T,U}$ is a subgroup of this $GL(4)$
as already mentioned above
(see \cite{CaLuMo} for an explicit embedding), 
$M\cdot N$ is therefore an  $O(2,2,{\bf Z})_{T,U}$ invariant.
Likewise $M' \cdot N'$ and $M'' \cdot N''$
are invariant under $O(2,2,{\bf Z})_{S,U}$ and $O(2,2,{\bf Z})_{S,T}$,
respectively.

When solving the three quadratic equations (\ref{QuadEq}) one encounters
the quantity
\be
D = \la M, M \ra \la N, N\ra - (M \cdot N)^2
\eq
and its transformed $D'$ and $D''$ under $S \leftrightarrow T$ 
and $S \leftrightarrow U$
as discriminants. $D$ is manifestly invariant under
$O(2,2,{\bf Z})_{T,U}$ but one can easily check that it is 
also invariant under the exchange transformations
$S \leftrightarrow T$ and 
$S \leftrightarrow U$ and therefore under the full triality group:
$D= D' = D''$. Note that it was argued in \cite{CCLMR} that
no quadratic invariant of the full triality group can be 
constructed out of $M_I,N^I$.
This is no contradiction to the result here because the invariant
$D$ is quartic.

In order to obtain
extremized moduli with a positive real part, one
has to demand that $D > 0$. For $D \leq 0$ the moduli
become purely imaginary. This does only make sense if 
they are imaginary rational, $S,T,U \in i {\bf Q}$. These
values are cusps of the corresponding $SL(2, {\bf Z})$ and
are thus equivalent to $S=T=U= \infty$ which, of course, is not an
interesting solution. For $D>0$ the explicit solution of (\ref{QuadEq}) 
is
\begin{eqnarray}
S &=& i \frac{M \cdot N}{\la N,N \ra} 
+ \sqrt{ \frac{\la M ,M \ra }{\la N, N \ra}
- \frac{ (M \cdot N)^2 }{ \la N, N \ra^2 } } \;\;, \nonumber \\
T &=& i \frac{M' \cdot N'}{\la N',N' \ra} 
+ \sqrt{ \frac{\la M' ,M' \ra }{\la N', N' \ra}
- \frac{ (M' \cdot N')^2 }{ \la N', N' \ra^2 } } \;\;,  \nonumber \\
U &=& i \frac{M'' \cdot N''}{\la N'',N'' \ra} 
+ \sqrt{ \frac{\la M'' ,M'' \ra }{\la N'', N'' \ra}
- \frac{ (M'' \cdot N'')^2 }{ \la N'', N'' \ra^2 } } \;\;. 
\label{Solution}
\end{eqnarray}

The transformation properties of the solutions under triality transformations
are manifest: the solution for $S$ is invariant under the $T$ duality group
$O(2,2,{\bf Z})_{T,U}$, whereas $T$ and $U$ are exchanged under
$T \leftrightarrow U$ 
(which acts as $M' \leftrightarrow M'', \; N' \leftrightarrow N''$
on the quantum numbers) and transform fractional linearly under 
$SL(2,{\bf Z})_T$ and $SL(2,{\bf Z})_U$. On the other hand
the solutions for $S$ and $T$ and $S$ and $U$ are 
exchanged by ${\bf Z}_2^{S-T}$ and ${\bf Z}_2^{S-U}$
respectively. 
And the solutions for 
$T$ and $U$ are invariant under the groups 
$O(2,2,{\bf Z})_{S,U}$, $O(2,2,{\bf Z})_{S,T}$.

\section{The entropy \label{classical}}
\setcounter{equation}{0}

It is now straightforward to compute the entropy with
the result 
\be
{\cal S} / \pi = M^2_{ADM} = 
\sqrt{ \la M, M \ra \la N, N \ra - (M \cdot N)^2 }
= \la N, N \ra \, \Re S \;\;.
\label{ClassEntr}
\eq
This agrees with the result obtained in \cite{KalShmWon} 
by considering so called double extreme solutions, in which
the moduli take the same value at infinity and on the horizon
(and are therefore constant in between). 
As shown above the classical entropy is invariant under the full
triality. This had to be expected because the classical BPS spectrum
from which it is computed has this property.

{From} (\ref{ClassEntr}) one can easily read off that the entropy
vanishes for certain classes of black holes. 

The first class consists of black holes with only four non--vanishing
quantum numbers, including the cases of purely electric ($N=0$) and
purely magnetic ($M=0$) black holes. In fact, it is a well known
property of extremal $N=2$ black holes that only dyonic ones
can have a non--vanishing event horizon and entropy \cite{Tse}.
By triality we know that the entropy must also vanish if
$M'=0$ or $N'=0$ or $M'=0$ or $N''=0$. 

The second class of black holes with vanishing entropy has
eight non--vanishing quantum number which are, however, not completely 
independent from one another, with the effect that the quantum 
numbers can be expressed in terms of momentum quantum numbers $m_i$
($i=1,2$) 
and winding quantum numbers $n^i$ of strings around the two--torus
as
\be
M_I/p=(m_2,-n_2,n_1,-m_1), \;\;\;N^I/q=(-n_2,m_2,-m_1,n_1) \;\;.
\eq
This implies that 
\beqa
\la M,  M \ra  &\equiv& - 2 p^2 n^T m  \;\;, \nonumber\\ 
\la N, N \ra &\equiv&  - 2 q^2 n^T m   \;\;, \nonumber\\ 
M \cdot N &\equiv&   - 2 pq n^T m \;\;, \label{Hshort} 
\eeqa
and therefore ${\cal S}=0$. The corresponding BPS states 
are those which can be embedded into a heterotic $D=4, N=4$
model as short multiplets \cite{CCLMR}, implying that only those states
which are intermediate from the $N=4$ point of view contribute
to the entropy. By triality the entropy also vanishes for those
black holes, where the vectors $M',N'$ and $M'',N''$ take
the form (\ref{Hshort}). In $D=4,N=4$ theories the exchange symmetries
$S \leftrightarrow T$ and $S \leftrightarrow U$ 
do not act as a self--duality but map heterotic to $IIA$ and
$IIB$ strings, respectively \cite{DufLiuRah}, \cite{CCLMR}. 
In particular they map short multiplets
of the heterotic theory to multiplets which are intermediate in
the heterotic but short in the $IIA$ or $IIB$ theory \cite{BehDor},
\cite{CCLMR}. Therefore,
the triality rotated $N=2$ black holes with vanishing entropies
are those which come from short multiplets of the $N=4$
$IIA$ or $IIB$ theory.

Finally we would like to recall the entropy formula for heterotic
$N=4$ black holes 
\cite{D4N4Entr}
\be
{\cal S} / \pi = \sqrt{ P^2 Q^2 - (Q \cdot P)^2 } \;\;, 
\label{n4entro}
\eq 
where the $Q$ denote 28 electric charges and where the $P$ denote
28 magnetic charges which lay in the (6,22) Narain lattice.  $O(6,22,{\bf Z})$
invariance then restricts the entropy to be of the form (\ref{n4entro}).
Comparing to (\ref{ClassEntr}), it is evident that the $N=2$ formula
should result from a suitable truncation.
Short $N=4$ multiplets are characterized by $P$ and $Q$ being
parallel. Thus it is evident that only intermediate $N=4$ multiplets
contribute to the entropy.

\section{Quantum Corrections \label{quantum}}  
\setcounter{equation}{0}

In $N=2$ supergravity coupled to vector multiplets, there are both
perturbative and non--perturbative quantum corrections.
The full prepotential of the $S$-$T$-$U$ model takes the
form \cite{CDFP}, \cite{DKLL}, \cite{AFGNT}
\be
F(S,T,U) =  - STU + f(T,U) + f^{(NP)} (e^{-2 \pi S}, T, U).
\eq
Note that the perturbative correction $f(T,U)$ does not
depend on the dilaton, reflecting the fact that perturbative
corrections can occur at the one-loop level, only. 

Let us first neglect non--perturbative corrections and consider
the one-loop effects.
Obviously,
the full perturbative theory with prepotential 
$F^{pert}(S,T,U) = -STU + f(T,U)$ is not triality symmetric.
Moreover $T$--duality is still a symmetry, but the
symplectic transformations get modified such that the
perturbative $T$ duality group is no longer $O(2,2,{\bf Z})_{T,U}$.

In order to obtain 
a convenient description of the perturbative $T$ duality group, 
a transformation from the symplectic section $(X^I, i \p_I F)$
(where $(X^I) = e^{K/2} (1,iS,iT,iU)$),
defined in terms of the prepotential, to the section $(P^I, i Q_I)$
introduced earlier by means of the symplectic transformation
$P^1 = - i F_1 (=TU)$, $iQ_1 = X^1 (= iS)$ is required \cite{CDFP},
\cite{DKLL}.

In the new parametrization, the one-loop $T$ duality transformations take
the form
\be
\Gamma^{Tree} =
\left( \begin{array}{cc} 
U & 0 \\ 0 & U^{T,-1} \\
\end{array} \right)
\longrightarrow
\Gamma^{Pert} =
\left( \begin{array}{cc} 
U & 0 \\ W & U^{T,-1} \\
\end{array} \right) 
=
\left( \begin{array}{cc} 
U & 0 \\ 0 & U^{T,-1} \\
\end{array} \right) 
\left( \begin{array}{cc}
1 & 0 \\ \Lambda & 1 \\
\end{array} \right) \;\;,
\eq
where $\Lambda$ is a symmetric integral matrix which encodes the quantum
corrections.

The one-loop transformation rule of symplectic quantum numbers 
implied by the general formula $(N , -M) \rightarrow (N,-M) \Gamma^T$
is
\be
M \rightarrow U^{T,-1} M - W N,\;\;\;N \rightarrow U N \;\;.
\eq

Explicit generators of the perturbative $T$ duality group have been
given in \cite{DKLL}, \cite{AFGNT}, \cite{CaLuMo}. 
The most obvious effect is that
the dilaton $S$ is no longer invariant under perturbative $T$ duality
but is shifted by a function of $T$ and $U$. One can however
define an invariant dilaton $S_{invar}$ by adding a
suitable function of $T$ and $U$ to the dilaton $S$ \cite{DKLL}. 
Note that the invariant dilaton is not a special $N=2$ coordinate.

The transformation law of the classical entropy under one-loop
$T$--duality transformations can be easily worked out and reads
\be
{\cal S} = \pi \sqrt{D} \rightarrow {\wt{\cal S}} = \pi \sqrt{ \wt{D} } \;\;,
\eq
where
\beqa
\wt{D} - D &=& ( \la WN, WN \ra - 2 \la WN, U^{T,-1}M \ra ) \la N, N \ra
\nonumber\\
&& + 2 (M \cdot N) (WN \cdot UN) - ( WN \cdot  UN )^2 \;\;.
\eeqa
Thus, the tree level expression for the entropy 
is not invariant under
one-loop $T$ duality.
Whereas the breaking of the full triality symmetry
by loop corrections is no surprise, one expects $T$ duality to be 
true at the perturbative level and thus there must exist a modification
of the entropy formula. 
One can also check that the left and right hand sides of the classical
solutions (\ref{Solution}) for the moduli on the horizon transform
differently under perturbative $T$ duality. This implies that
the solutions must be modified at the one-loop level, too.
We expect that invariants of the perturbative $T$ duality
group will play a crucial role. Note that $\la N, N \ra$
is such an invariant whereas $\la M , M \ra$, $M\cdot N$ and
$D$ are not.

A natural candidate for the one-loop entropy is
found by observing that the tree level entropy can be written
as
\be
{\cal S}_{tree} = \pi  \la N, N \ra \Re S \;\;.
\eq
Now, $\la N , N \ra$ is invariant under one-loop $T$ duality, 
and
it is well known that one can make the dilaton invariant by adding
a suitable function of $T$ and $U$, yielding the so--called 
invariant dilaton $S_{invar}$ \cite{DKLL}. 
This motivates us to conjecture that
the one-loop entropy is given by
\be
{\cal S}_{1-loop} =   \pi  \la N , N \ra \Re S_{invar} \;\;.
\eq
The conjecture is also compatible with the known perturbative structure
of gravitational threshold corrections. Since at the tree level
only states, which are intermediate from the $N=4$ point of view 
contribute to the entropy we expect that at one--loop all contributions
of such states go into the invariant dilaton.

Let us finally recall what is known about non--perturbative effects.
It is firmly established that the heterotic $S$-$T$-$U$ model
is dual to the $IIA$ compactification on the Calabi--Yau threefold
${\bf P}_{1,1,2,8,12}(24)$ \cite{KacVaf}, \cite{KleLerMay}, 
\cite{KapLouThe}, \cite{AGNT}, \cite{Cur}, \cite{CCLM}. 
This implies that the full non--perturbative prepotential
is given by the classical prepotential of the dual $IIA$ model.
Symplectic matrices corresponding to the true quantum symmetries
can be computed by studying the monodromy properties
of the prepotential around its singular loci on the
moduli space of the threefold.

Both the structure of the conifold locus of the threefold
and an analysis of the perturbative monodromies
of the heterotic theory indicate the following:
$T$ duality transformations
corresponding to Weyl transformations of generically
Higgsed non--Abelian gauge groups
(like for instance the $T \leftrightarrow U$ exchange,
which is the Coxeter twist of the $SU(2)$ gauge group
unbroken at $T=U$)
are replaced by two non--perturbative quantum monodromies
caused by dyons that become massless. 
This is the stringy generalization of the Seiberg-Witten
solution of the pure $N=2$ $SU(2)$ super Yang--Mills theory.

Since the Calabi--Yau moduli space contains more singular loci,
there is space for other non--perturbative effects
as well. There exists, for example, the so--called
strong coupling locus \cite{KleMay}, which is fixed under the 
exchange transformation $S \leftrightarrow T$ \cite{KleLerMay},
\cite{CCLMR}, \cite{CCLM}. Thus, the 
$S \leftrightarrow T$ exchange
is, although not a symmetry at the perturbative 
level, a symmetry of the full non--perturbative theory.

Since the group structure of symmetries is 
strongly modified when going from the perturbative to
the non--perturbative level, one expects that the 
entropy formula is also further modified. A better
knowledge of the monodromy group and of its invariants
should be useful for dealing with this question.

%\newpage
\vspace{1cm}

{\bf Acknowledgement}\\

The work of T.M is supported by DFG. He thanks K. Behrndt for
many useful discussions.

While preparing this manuscript we received the preprint \cite{BKRSW}, 
which contains partially overlapping independent work.

\end{document}